\documentclass[12pt,a4paper]{article}
\usepackage{graphicx}
\begin{document}
\textwidth=135mm
 \textheight=200mm
\begin{center}
{\bfseries Prompt photon $A_{N}$ with the PHENIX MPC-EX detector}
\vskip 5mm
J. Lajoie$^{\dag}$ for the PHENIX Collaboration
\vskip 5mm
{\small {\it $^\dag$ Iowa State University, Dept. of Physics and Astronomy, Ames, Iowa 50021}} \\
\end{center}
\vskip 5mm
\centerline{\bf Abstract}
The PHENIX MPC-EX detector is a Si-W preshower extension to the existing Muon Piston Calorimeters (MPC).  
%The MPC-EX will consist of eight layers of alternating W absorber and Si "minipad" sensors and will be installed and ready for physics in time for RHIC Run-15.  Covering the pseudorapidity range $3.1 <|\eta| < 3.8$, the MPC-EX and MPC access high-x partons in the projectile nucleon 
%(and low-x partons in the target nucleon)
 %in transversely polarized proton-proton collisions at 200 GeV.  
%With the addition of the MPC-EX, the neutral pion reconstruction range extends to energies $>$ 80 GeV, a factor of four improvement over current capabilities.  
%Not only will the MPC-EX strengthen PHENIX's existing forward
%$\pi^{0}$ and jet measurements, it also provides the necessary $\pi^{0}$ rejection to make a prompt photon measurement feasible.  
%With this $\pi^{0}$ rejection, prompt photon yields for $p_T >$ 3 GeV/c can be statistically extracted using a double ratio method. 
The combined detectors will make possible a measurement of the prompt photon single spin asymmetry $A_N$ in 200GeV transversely polarized
$p$+$p$ collisions, 
which will help elucidate the correlation of the transverse motion of valence partons in the proton with the proton spin.  

\vskip 10mm
\section{\label{sec:intro}Introduction}
The large transverse single-spin asymmetries observed in polarized $p$+$p$  collisions at RHIC
\cite{BRAHMS2008,STAR2008}
are believed to be related to a combination of initial and finall state effects that originate primarily in
the valence region of the projectile nucleon.
% (the Sivers or transversity distributions in the
%TMD approach, or parton correlations in a collinear factorized framework). 
While data
in semi-inclusive deep-inelastic scattering has been used to constrain these effects \cite{ANSELMINO_1}, the
situation is more complicated in $p$+$p$ collisions due to the presence of both strong initial and
final-state corrections arising from the soft exchange of gluons.

Existing measurements at forward rapidity in transversely polarized $p$+$p$ collisions are limited 
to hadronic observables, and therefore sample a number of different partonic processes. In contrast, 
direct photon production at forward rapidity is dominated by 
the scattering of a valence quark from the polarized projectile off a low-$x$ gluon in the unpolarzied
proton. 
%Prompt photons at forward rapidities are roughly an equal mix of direct photons and photons
%from the fragmentation of the scattered valence quark. 
Theoretical studies indicate that the 
contribution from transversity and a polarized fragmentation function is small \cite{KANG2012_1}, 
so that the single spin asymmetry for fragmentation photons largely carries the same information 
about the initial state of the polarized nucleon and reinforces the asymmetry from direct photons. 
Fragmentation photons can be difficult to disentangle from direct photons, so they are 
commonly combined and referred to as ``prompt" photons.   

%The identification of prompt photons is very challenging experimentally. 
The prompt photon signal 
must be extracted from a large background of photons from hadronic decays, primarily from $\pi^{0}$
and $\eta$ mesons. 
%The Muon Piston Calorimeter (MPC) Extension, or MPC-EX, is a Si-W preshower detector
The MPC-EX is a Si-W preshower detector
that will be installed in front of the existing PHENIX MPC’s. The MPC's are lead-tungstate 
electromagntic calorimeters covering rapidities $3.1 <|\eta| < 3.8$.  The new detector will consist of
eight layers of Si “minipad” sensors interleaved with W absorber and enable the
identification and reconstruction of  $\pi^{0}$ mesons at energies up to $>80$ GeV.
The capability of the MPC-EX to reconstruct and reject $\pi^{0}$ mesons (as well as other
hadronic sources of photons) at very high energies is essential to separate prompt
photons from other sources of photons. 

\section{The MPC-EX Preshower Detector}

%The location and design of the MPC-EX are shown in Figure~\ref{fig:phenix}. 
The MPC-EX will
be located approximately two meters from the interaction vertex, inside the muon magnet pistons. 
The MPC-EX itself will consist of eight measurement layers, each layer separated by a 2~mm thick W absorber. The active 
area of each layer will consist of 500$\mu m$ thick Si "minipad" 6.2 x 6.2 cm$^2$ sensors. The sensor itself is divided 
into 1.8~mm by 15mm minipads to provide 128 separate channels per sensor. The orientation of the sensors will alternate by $90^o$ 
in successive layers. The readout electronics will consist of two SVX4.2b
ASICs per module, and a capacitive split will provide both a high gain and low gain channel for each minipad, for a total or 256
channels per module. The modules will be installed on an electronics carrier board that will provide readout and bias 
voltage for the individual sensors. The carrier board itself will be bonded to the W absorber for each layer. 
%The design of the MPC-EX is highly modular, allowing for simplified fabrication and maintainence.  

Clusters from high momentum $\pi^{0}$ decays ($E>20$ GeV) merge in the MPC and are indistinguishable from a single 
high energy photon.  With the MPC-EX in place, the resolution of the individual minipad elements generates a 
highly detailed picture of the early development of the electromagnetic shower of each photon. Distinct centers of gravity can 
be identified for each photon shower and a four vector reconstructed from the MPC-EX shower and energy, and the total energy
deposited in the MPC. These four vectors can be combined to reconstruct an invariant mass for the combined shower. 
This technique will allow the measurement of the $\pi^{0}$ and $\eta$ meson production nearly out to the kinematic limit in 200~GeV $p$+$p$ collisions. 

\section{Prompt Photon $A_N$}
Extensive simulations have been undertaken to demonstrate the ability to extract a prompt photon signal using the MPC-EX. Approximately 870M 
%{\sc Pythia}\cite{PYTHIA} 
{\sc Pythia} 
events were generated that included all minimum bias processes as well as the production of direct and fragmentation photons. 
%In this was the data sample
%mimics a ``true" data sample, with the signal photons buried within background events. 
The event sample was put through a full simulation
of the PHENIX detector, including the MPC-EX. The data was digitized with the response of the detectors and then reconstructed.  A set of basic cuts to 
remove background direct photons from $\pi^{0}$ and charged hadron interactions were applied.  Because the ratio of prompt photons to photons from $\pi^{0}$
decays improves as a function of transverse momentum we examined all candidates with a transverse momentum $p_T>3$GeV/c.  With these cuts the reconstruction 
efficiency for prompt photons was 31\%, while the reconstruction efficiency for photons from $\pi^{0}$ mesons was
only 2.9\%, indicating an improvement in the prompt photon/$\pi^{0}$ ratio of more than a factor of ten.  
%Summing over all background sources, the S/B for prompt photons was 0.34 for $p_T>3$ GeV/$c$.  
%Of the true prompt photons in the sample, 57\% were direct photons, with the remainder
%being fragmentation photons. 

In a direct photon measurement with the MPC-EX detector, the signal of prompt photons ($S$) and
the background of hadron-decay photons ($B$) are mixed.
Simulations show $S/B\approx 0.4-0.5$ at $p_T>3$ GeV/$c$.
%According to the simulations described above, with a cut of $p_T> 3$ GeV/c
%the value of $r$ typically is $\approx 0.4-0.5 $.
In measurements of a prompt photon $A_N$ in 200~GeV $p$+$p$ collisions, the background photon events will carry a non-zero asymmetry, such as
 from $\pi^0$ or $\eta$ decay photons. The background asymmetry can be subtracted based on measurements of the $\pi^{0}$ and $\eta$ asymmetries. 

%The signal asymmetry ($A_S$) can be extracted from the measured asymmetry ($A_{meas}$ )
%and the independently measured background asymmetry ($A_B$) according to:

%\begin{equation}
%A_S = (1+{\frac{1}{r}}) A_{meas} - \frac{1}{r} A_B.
%\end{equation}

%\begin{equation}
%(\delta A_S)^2 = (1+{\frac{1}{r}})^2 (\delta A_{meas})^2  +  (\frac{1}{r})^2 (\delta A_B)^2.
%\end{equation}

\begin{figure}
\centering
\includegraphics[width=3.3in]{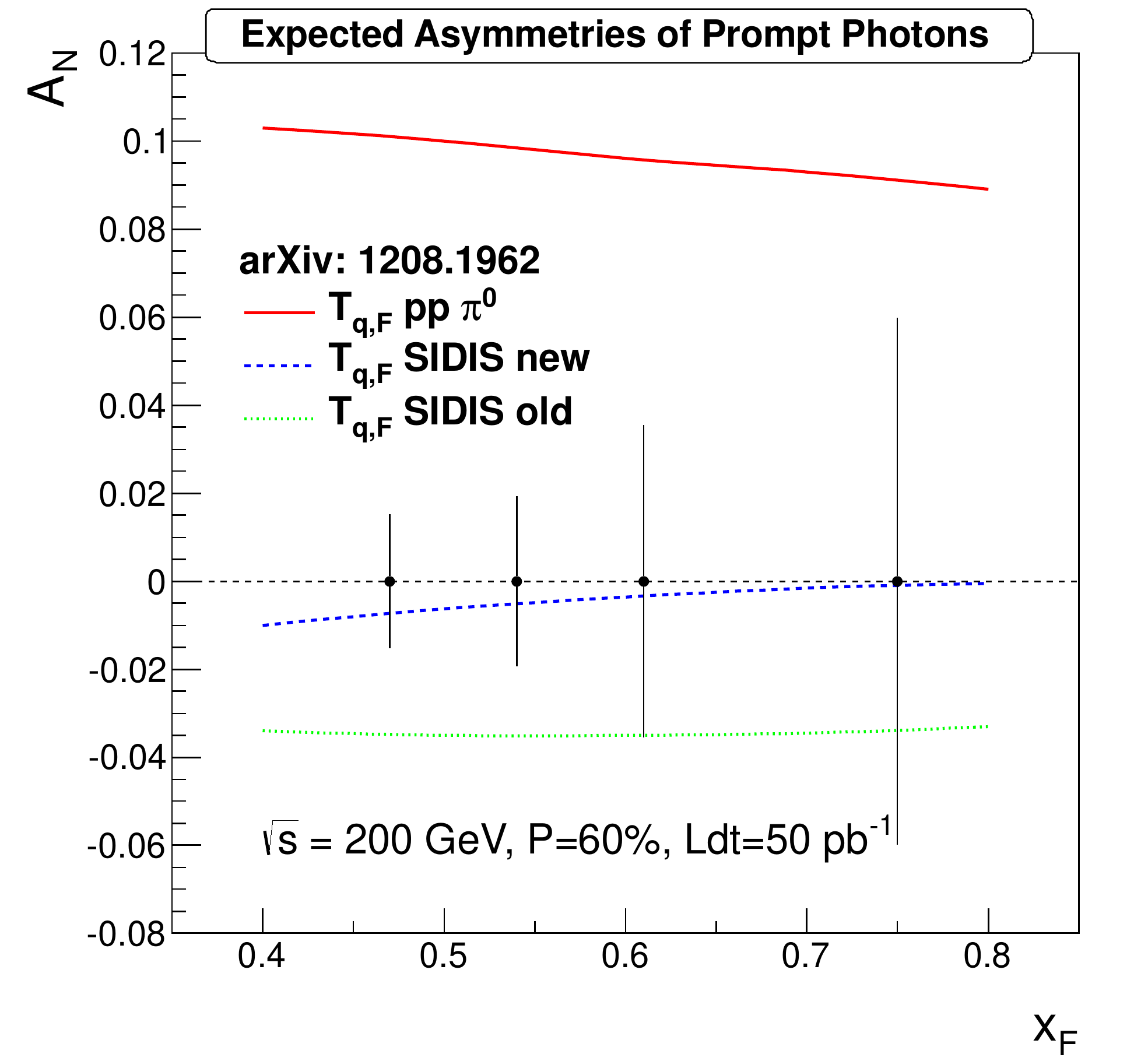}
\caption{\label{fig:AN}Projected sensitivity for the prompt photon single spin asymmetry with the MPC-EX assuming an integrated luminsoity of 
50 pb$^{-1}$ and 60\% beam polarization. The sensitivities are shown compared to calculations in the collinear factorized approach \cite{KANG2012_1,KANG2011_1} 
using a direct extraction of the quark-gluon correlation function from polarized $p$+$p$ data (upper solid curve), compared to the correlation function derived from SIDIS extractions (lower dotted and dashed curves).}
\end{figure}

Assuming a sampled luminosity of 50 pb$^{-1}$  and a cut of $p_T > 3$ GeV/c,
0.75 million photon events will be observed in the MPC-EX. The simulated events are split into bins  that  corresponding to
central values of  ($p_T$,  $x_F$, number of events) as:  (3.2,  0.47,  400k),  (3.6, 0.54, 250k), (4.0,  0.61, 75k) and
(4.4, 0.75, 25k). We assume that MPC-EX can independently measure the asymmetry of mesons the $\pi^0$ and $\eta$ mesons 
to a precision at least a factor of two better than the inclusive photon asymmetry in each bin.  The estimated precision
of  the prompt photon $A_N$ is shown in Figure \ref{fig:AN}. 
%with theory predictions for the prompt photon $A_N$ of
%Kang {\it et al.}\cite{KANG2012_1,KANG2011_1}.

\section{Conclusion}

The MPC-EX is a novel Si-W preshower detector that will enable the measurement of prompt photons as forward rapidities 
in transversely polarized $p$+$p$ collsions at $\sqrt{s}=200$GeV. The detector is currently under construction, and a test run with partial 
instrumentation is planned for fall 2013, (RHIC Run-14) with complete installtion in both PHENIX arms for fall 2014 (RHIC Run-15).  
Measurement of the prompt photon single spin asymmetry in transversely polarized $p$+$p$ collisions will help elucidate the origin of single spin asymmetries in hadronic collisions.

%\begin{figure}
%\centering
%\includegraphics[width=5.0in]{figure1.pdf}
%\caption{\label{fig:phenix}The PHENIX detector (upper right), showing the location of the existing Muon Piston Calorimeters inside the
%muon magnet piston. The MPC-EX (upper left) will consist of eight measurement layers of absorber, sensors and readout.  The "minipad" sensors 
%themselves (lower right) will consist of a readout card bonded to a Si sensor. The orientation of the long direction of the minipads will alternate between
%layers. }
%\end{figure}

\end{document}